\documentclass[aps,3p]{elsarticle}
\usepackage{epsfig}
\usepackage{xcolor}
\usepackage{epsfig,graphicx,float,pst-all}
\usepackage{mathrsfs}
\usepackage[utf8]{inputenc}
\usepackage{url}
\usepackage[caption = false]{subfig}
\usepackage[normalem]{ulem}
\usepackage{multirow}
\begin{document}
	\newcommand{\gs}[1]{\textcolor{red}{\it #1}}
	\newcommand{\gsout}[1]{\textcolor{red}{\sout{#1}}}
	\newcommand{\lf}[1]{\textcolor{blue}{\it #1}}
	\newcommand{\lfout}[1]{\textcolor{blue}{\sout{#1}}}
	\begin{frontmatter}
		\title{Pairing enhancement in a two-neutron transfer process through the continuum}

		
		\author{G. Singh}
		\ead{g.singh@unipd.it}
		\author{L. Fortunato}
		\author{A. Vitturi}
		\address{Dipartimento di Fisica e Astronomia ``G.Galilei'', Università degli Studi di Padova, via Marzolo 8, Padova, I-35131, Italy}
		\address{INFN-Sezione di Padova, via Marzolo 8, Padova, I-35131, Italy}
		
		
		\date{\today}

		\begin{abstract}
			Enhancement, due to constructive interference through the many possible reaction channels, occurs in two-neutron transfer reactions from a bound system $A$ to a weakly-bound system ($A$+2), whenever the intermediate system is unbound and consists of a continuum with a resonant state. We elucidate this phenomenon in the case of $^6$He, modeled as two neutrons in the orbitals of an intermediate $^5$He nucleus, comparing the realistic case of an unbound ($A$+1) system with an artificially bound case. A properly modeled continuum in these calculations is found to be crucial and leads to significant enhancement due to pairing correlations.
		\end{abstract}

	\end{frontmatter}

	\section {Introduction}
	
	Direct transfer reactions have proven to be convenient tools to study light exotic nuclei, since the reaction mechanisms are simple enough to probe the structure of participating nuclei with sufficient accuracy. 
	Two-neutron transfer reactions, in particular, enable a better comprehension of the halo phenomenon \cite{JRF04RMP,SSC22PRC,BKT20PRL,FCH20CP,SC14NPA,MDS21NPA,SSC16PRC} in two-neutron halos by providing considerable information about the correlations between valence neutrons \cite{TAB08PRL}. They have been especially important in trying to explain mass gaps at $A$ = 5, 8 through dineutron captures on $^4$He and $^6$He. In fact, two-neutron captures or transfers to form $^6$He are also known to have a considerable effect on $r$-process seed nuclei production in the extremely neutron rich environment at low temperatures \cite{BGM06PRC,GHT95PRC}.
	
	{An {imbalanced} neutron to proton ratio in exotic nuclei forces the Fermi surface to approach the particle threshold and develops weak separation energies of valence nucleon(s). As a consequence, the existence of diffused halos and the possibility of Cooper pairs scattering into the continuum can affect the character of short-range correlations \cite{OV01RPP}. As the particles gain energy and excite to higher energy levels due to the introduction of additional correlations to simple mean field models, the role and inclusion of the continuum then becomes indispensable \cite{MN06NPA,VP10NPA,MMV21PRC,VMH15AIP,HOM22PPNP,UDN12PRC}, especially if breakup or transfer channels are dominant. However, the wave functions of these continuum states are not square integrable and hence, one loses the convergence and orthonormality conditions in most cases, making the continuum considerations a non-trivial task for most investigations.}
	

	Furthermore, the couplings amongst such continuous positive energy states and pairing correlations strongly enhance two particle or cluster transfer reactions \cite{OV01RPP,VMH15AIP}. This pairing enhancement is thought to originate from `coherent interferences' of different paths through the intermediate states in the ($A$ + 1) nuclei {that arise due to the correlations in the initial and final state wave functions. These correlated initial and final state wave functions provide the necessary microscopic and nuclear structure information to obtain two particle transfer amplitudes, giving the weight of each of the two-step paths, which in turn can be accessed from single particle transfer form factors.} For weakly bound two-neutron halo nuclei, this would mean strengthening of the transfer process {to the ground state} {of the final nuclear product} via the continuum (unbound) states of the intermediate nucleus.  
	
	A quantification of such a process is yet to be done taking fully into account these coupling effects in the continuum. As an intuitive guess, one would expect to have a decreasing probability contribution for the pairs in higher states of the continuum, unless of course, a resonance is encountered. Therefore, a detailed study of coupled channels and sequential transfer effects needs to be undertaken {for weakly bound nuclei} taking into thorough consideration the dynamical interplay of the continuum \cite{TAB08PRL,MMV21PRC,VMH15AIP,MD21PLB} through the intermediary nucleus and compare them with the bound state couplings (if any) to understand the reaction mechanisms properly. We try here such a case and study the effect of pairing through the {photography of the ground state} and the continuum of $^5$He in a two-neutron transfer reaction for the formation of the simplest two-neutron halo nucleus $^6$He. We consider scenarios where the $^5$He intermediary nucleus is unbound (as is the natural case) and also bound hypothetically with the bound ground state at 1\,MeV and 0.1\,MeV to form two different cases for comparisons. 
	
	The inclusion of the continuum, as mentioned, is difficult due to the continuous nature of the energy states when the solution of the Schr\"{o}dinger equation goes above zero, i.e., $E > 0$. A method to deal with this issue is to discretize it into a finite number of states for meaningful calculations. We achieve this through the pseudostate method of discretization by expanding the continuum states in a transformed harmonic oscillator (THO) basis \cite{PMA01PRA,RAG05PRC,KAG05PRC}, although other methods can be equally efficient \cite{MPV16JPG}. THO maintains the lucidity of the harmonic oscillator functions, converting, however, their Gaussian asymptotic behavior to a better suited exponential mode \cite{CRA13PRC}. For our purpose, we use a maximum of 8 states in total
	\footnote{We could easily increase this number, but we achieved sufficient convergence for all practical purposes and wished to maintain it as small as possible to avoid an explosion of the basis states in the two-neutron transfer calculations, which lead to an enormous computational cost due to the couplings. Nevertheless, it should not be too small either.}, and ensure the orthogonality of our channels to avoid any consideration of a simultaneous transfer process \cite{BWBook}.
	
	The generation of these wave functions involved fixing the parameters of the THO basis at optimal values. An advantage of using the THO basis is that one can control the density of pseudostates representing the continuum near the threshold by modifying the $\gamma/b$ ratio of the basis (where $\gamma$ is related to the transition radius and $b$ is the oscillator strength) \cite{KAG05PRC,Casal18PRC,LMA10PRC}. A smaller ratio results in a larger density of states near the continuum {threshold}, a desirable trait for weakly bound nuclei since they have only one or two bound states. More details about the THO basis and these parameters can be found in Refs. \cite{SSC22PRC,CRA13PRC,Casal18PRC,LMA10PRC,LMA12PRC,CSF20PRC}. In the present study, we fixed the values of $\gamma/b$ at 2.0, 1.8 and 1.67\,fm$^{-1/2}$ for the three cases of $^5$He nucleus, i.e., bound with $S_n$ = 1\,MeV, bound with $S_n$ = 0.1\,MeV and unbound, respectively. These values were chosen for each case so as to obtain the optimum spatial and density representation of the discretized continuum for a given energy range \cite{MAG09PRC,LMA10PRC}.
	
	It is noteworthy that in the unbound case, the third state in our spectrum generation was closest to the experimentally observed resonance in $^5$He and was fixed for our purpose at 0.69\,MeV above zero. 
	For the bound cases, we adjusted the potential well of $^5$He so as to reproduce this state at values of 1\,MeV and 0.1\,MeV below the particle emission threshold, resulting in two bound states with different separation energies.
	
	\section{One-neutron transfer}
	
	We contemplate a sequential two neutron transfer for the formation of $^6$He via the reaction $^{18}$O($^4$He,$^6$He)$^{16}$O. The first step is a one neutron transfer reaction forming $^5$He 
	and leaving an intermediary $^{17}$O nucleus. For the sake of simplicity, we {do not consider any configuration mixing in our analysis yet and} take this $^{17}$O to be formed entirely of a $1d_{5/2}$ ground state bound by 4.1433\,MeV \cite{Wang17CPC}, while neglecting the contribution of the $2s_{1/2}$ state. Further, we only regard the transfer of the neutron to the $p_{3/2}$ state in the bound as well as continuum states of $^{5}$He. This is done with the aim of avoiding complications due to the mixing of different angular momenta.
	
	As mentioned above, the energies of the states obtained after discretization were adjusted by modifying the basis parameters to be as close to the resonance energy of $^5$He as possible for the continuum arrangement. As such, we were able to generate an energy level at 0.6912\,MeV in the continuum, which is close to the resonance value of the $p_{3/2}$ state at 0.789\,MeV \cite{FCS14PRC}. Notably, this was the third state in our basis. Therefore, we have a couple of states at lower energies and a bunch of states at higher energies, which together give a good representation of the entire range. 
	
	The beam energy in the lab frame $E_l$ used here was 21\,MeV, {which was just above the Coulomb barrier} and as such it was insufficient to populate the higher lying basis states due to a lower center of mass energy. Therefore, we also tried a lab energy of 100\,MeV that was sufficient to populate all the desired basis states in the continuum. This energy amounts to about 18\,MeV in the center of mass frame. Although this generates a larger number of states, we truncated the calculations at 6\,MeV due to negligible influence of higher lying states and the exponentially rising computational cost of the calculations.
	
	The form factors for this one neutron transfer reaction were then computed in the prior representation (cf. Chapter 5 of \cite{BWBook}) using a modified version of the Transformed Form Factors (TFF) code \cite{TFF}. {It was seen that the form factors were quite extended owing to the weakly bound nature of the nuclei, reflecting their non-locality and long range of overlap of the wave functions. This necessitates the inclusion of coupled channels calculations, taking into reckoning the continuum and larger widths of integration over the form factors to account for the non-orthogonality terms \cite{OV01RPP}.}
	
\begin{figure}[ht]
	\centering
	\includegraphics[trim={0 0 0 0},clip,width=9.9cm]{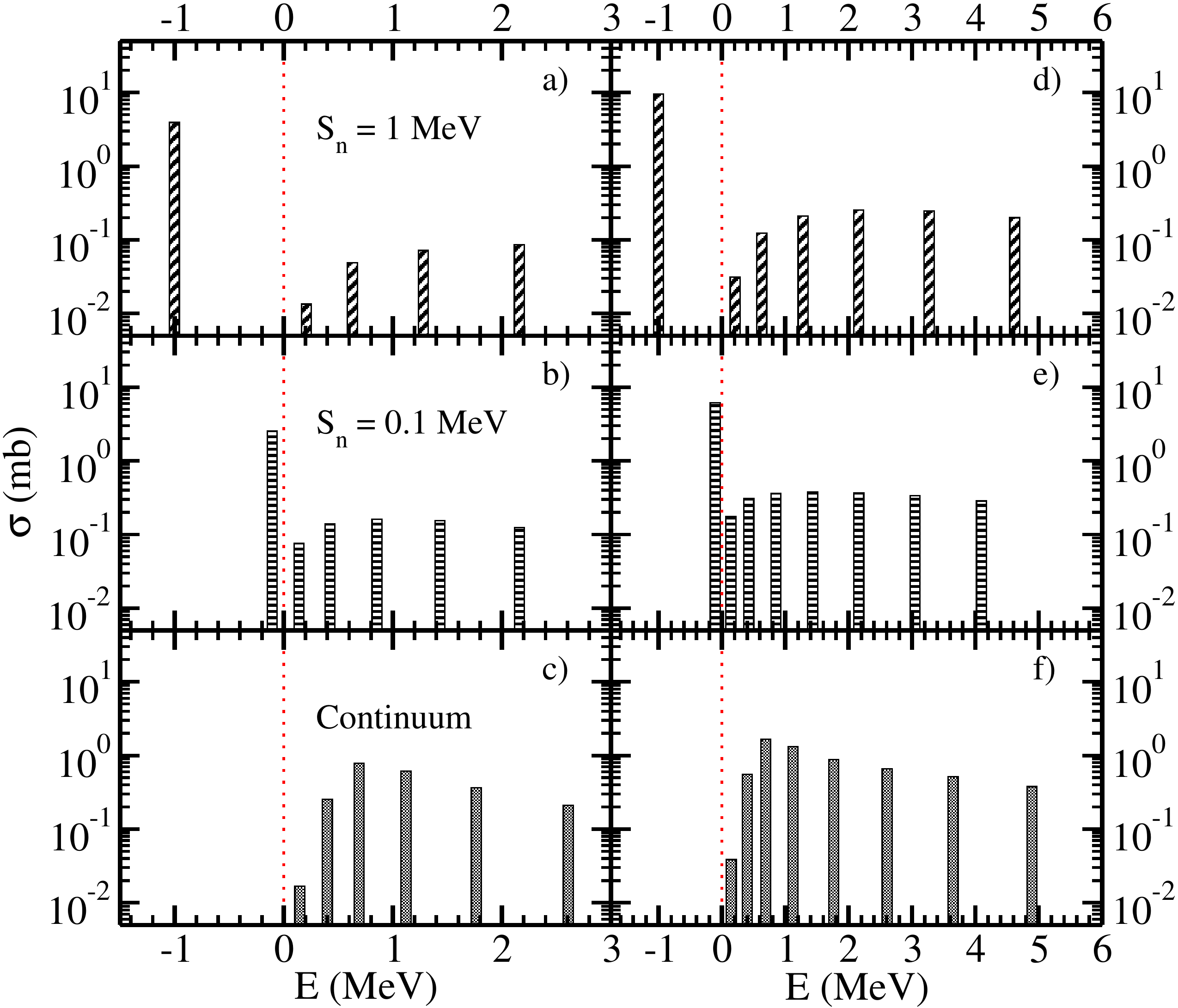}
	\caption{\label{fig: sigma_Sn}  Total cross-section for one-neutron transfer to form $^5$He in the three different scenarios. The beam energy for exhibits a), b) and c) was 21\,MeV, while that for d), e) and f) was 100\,MeV. As is evident, higher beam energy enables the population of higher lying states in the continuum.}
\end{figure}

	
With the states generated through the THO basis, the results for the total cross-sections for the one-neutron transfer reaction with contribution of each of the basis states at two different beam energies for all three cases is shown in Fig. \ref{fig: sigma_Sn}. The panels on the left shown as a), b) and c) correspond to the energy of the beam in the lab frame being 21\,MeV while those on the right marked d), e) and f) refer to the case of 100\,MeV beam energy. As stated above, when the beam energy was 21\,MeV, the continuum was truncated at 3\,MeV, while at 100\,MeV it was possible to include the basis states up to as much as 18\,MeV, but we truncate {and show} the results up to 6\,MeV {due to large computational costs.}
	
The clear dominance of the bound states to the contribution of the cross-sections is visible in the bound cases with $S_n$ = 1\,MeV and 0.1\,MeV. {Further, it is evident that the basis state closest to the resonance contributes the most for the continuum case.} It is noteworthy that traversing from panel a) to c), the contribution of the continuum starts to increase, another feature which is expected as the separation energy approaches zero and then changes sign. At 21\,MeV beam energy, for the two bound cases under consideration, we have continuum contributions of {0.12\,mb and 0.65\,mb,} while the total cross-section values stand at 4.73\,mb and 3.28\,mb, respectively. Meanwhile, for the continuum case, the total cross-section was computed at {2.08\,mb}. 
	
Ideally, increasing the lab energy should increase the cross-section. {At 100\,MeV beam energy, we find} by including 8 states, total one-neutron transfer cross-sections of {11.04\,mb} and {8.36\,mb} for $S_n$ = 1\,MeV and 0.1\,MeV, respectively. Naturally, the contributions of the continuum for these bound cases also arose accordingly to be at {1.10\,mb and 2.21\,mb}, respectively. At the same time the continuum case, with its continuum discretized using 8 states, gave a transfer cross-section of 5.95\,mb. 
	
\section{Two-neutron transfer}
	
Pairing interaction is vital when one considers the two-neutron transfer reaction. This involves the diagonalization of the Hamiltonian matrix (composed of an unperturbed part plus the pairing interaction) and obtaining the eigenvalues.
For the perturbative part, one actually needs a residual interaction between the continuum states of the two particles in $^6$He \cite{FCS14PRC}. Although density dependent interactions might be more appropriate \cite{BE91AP}, we chose a contact delta form, $-g\delta(\vec{r_1}-\vec{r_2})$ for the attractive pairing interaction as it simplifies the numerical implementation and involves only one parameter adjustment \cite{FCS14PRC,SFV16EPJ}.
	
{We adjusted the strength of this pairing interaction, $\Delta$, in order to keep the ground state energy of $^6$He at -0.975\,MeV \cite{Wang17CPC}. Table \ref{T1} {shows} the strength of the coupling constant $g$ (in MeV-fm$^{-3}$) required to lower the system energy to this desired value by adjusting the pairing interaction. The number of states $N$ representing the maximum number of pseudostates considered in the spectrum of $^5$He for the two beam energies were 6 and 8, respectively.} 
	
In case of the bound system with S$_n$ = 1\,MeV, a value of $g$ with an opposite sign was required 
	as the non-interacting two-neutron system has a larger binding energy than the state to be reproduced. Therefore, {we do not consider this case further and restrict our analysis to a bound case with $S_n$ = 0.1\,MeV and an unbound case of the continuum}. 
	
	
\begin{table*}[ht]
	\begin{center}
		\caption{\label{T1} {Table showing the two-neutron transfer cross-sections (in $\mu$b) corresponding to the coupling constant strength $g$ (in MeV-fm$^{-3}$) for each case under study at the two beam energies considered. The pairing interaction, $\Delta$ was varied to reproduce g.s. energy of $^6$He at -0.975\,MeV. The number of basis states contributing, $N$, for the bound and unbound case was 6 and 8, respectively. The unperturbed cross-sections $\sigma_{2n}$(u) and the ratio w.r.t. the pairing cross-sections are also indicated.}}
		\vspace{0.50cm}
		\begin{tabular}{|*{11}{c|}}
			\hline\hline
			Case & $\Delta$  & \multicolumn{4}{c|}{$E_l$=21\,MeV}  & \multicolumn{4}{c|}{$E_l$=100\,MeV} \\
			\cline{3-10}
				
			& (MeV) & $-g$ & $\sigma_{2n}$ & $\sigma_{2n}$ (u) & ${\sigma_{2n}}/\sigma_{2n}$(u) & $-g$ & $\sigma_{2n}$ & $\sigma_{2n}$ (u) & ${\sigma_{2n}}/\sigma_{2n}$(u) \\
			\hline
			$S_n$ = 0.1\,MeV & 0.775 & 1037 & 6.95 & 6.89 & 1.01 & 992 & 147 & 125 & 1.18 \\
			Continuum & 2.356 & 10430 & 0.94 & 0.51 & 1.84 & 7827 & 44 & 8.6 & 5.12 \\
				
			\hline\hline
		\end{tabular}
	\end{center}
\end{table*}
	
	\begin{figure}[htbp]
		\centering
		\includegraphics[trim={0 0 0 1cm},clip=true,width=8.3cm]{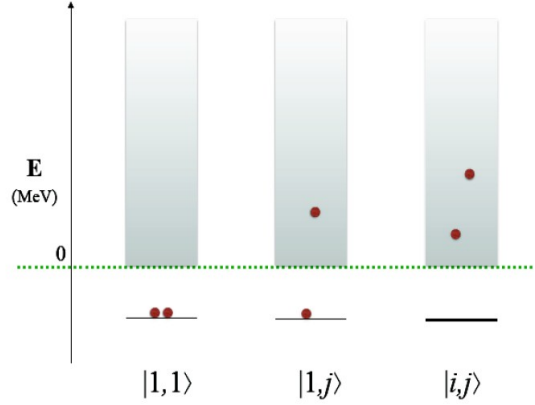}
		\caption{\label{fig: basis}  {Three possible configurations of a sequential two-neutron transfer process: both particles in the ground state, one particle in the continuum and two particles in the continuum.}}
	\end{figure}
	
	\begin{figure*}[htbp]
		\centering
		\includegraphics[trim={0 0 0 0},clip,width=15.3cm]{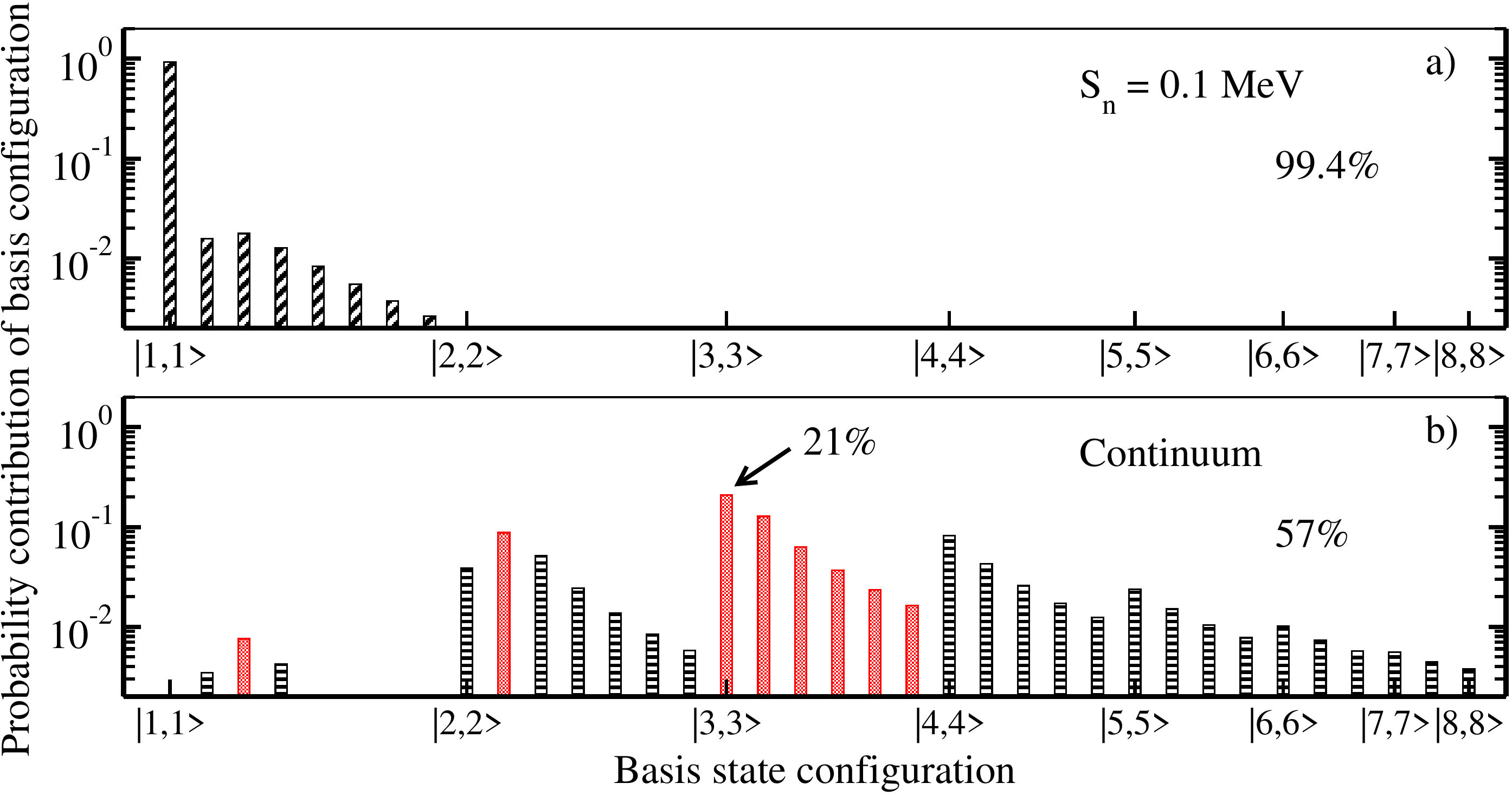}
		\caption{\label{fig: basis_contri}  {The cases of study for the sequential two-neutron transfer reaction $^{18}$O( $^4$He,$^6$He)$^{16}$O at 100\,MeV beam energy showing the probability contribution of each of the configurations through their respective bases. Seen in conjunction with Fig. \ref{fig: basis}, the pairing configurations $|i,i\rangle$ {give the highest} probability for particle occupation. For the bound system, the bound state contributes about entirely via the ground state configuration, while for the continuum case, each basis state couples to others giving considerable contributions. Further, in the lower panel, the dominant contribution of the basis state closest to the resonance energy is clearly visible and is highlighted by the (red) shaded bars.}}
	\end{figure*}
	
	Going from one particle transfer to subsequent, the number of configurations contributing to the transfer increase combinatorially (cf. Fig. 9 of Ref. \cite{OV01RPP}). Nevertheless, it is interesting and at the same time intuitive that the percentage contributions to the ground state of $^6$He by the different configurations of single particle basis states {arise from} those basis states that contribute the most to the one-neutron transfer cross-sections. Fig. \ref{fig: basis} {displays} the available scenarios for two particles when the entire basis is available for occupation. The {simplest} configuration {is found} when both particles occupy the ground state of the system, and as discussed later, {it} accounts for a substantial portion of occupation probability for the bound case. Secondly, one of the particles could lie in any of the other states while the first is in the ground state. Finally, we have the cases where both the particles are in any of the (excited) states, which may or may not be the same. Such a configuration, in general, has been denoted by $|i,j\rangle$ in our analysis ($i,j \in \{1,...,8\}$). 
	
	We mention again that although the number of two-particle configurations rises with the number of transferred particles, however, the pairing interaction ensures that most of the strength is collected into the ground state of the final nucleus \cite{OV01RPP}. 
	Therefore, one would expect the states with maximum pairing interaction {(the $|i,i\rangle$ configurations)} to contribute the most to the ground state. This is seen aptly in Fig. \ref{fig: basis_contri}, where we show the probability coefficients of each of the basis state configurations contributing to the system. {The $x$-axis shows the state that is generated through the basis and for brevity only the $|i,i\rangle$ configurations are written out. As we wanted to populate a higher number of states for proper comparisons, this calculation is shown only for 100\,MeV beam energy.}
	
	When both particles are in the same bound state, the configurations contribute as much as 93\%. A small contribution can also be seen when one particle is in the bound state, but the other is in any of the excited states of the continuum. These configurations would be $|1,j\rangle$ ($j\neq1$) and together with the $|1,1\rangle$ configuration, contribute about 99.4\% for the case when $^5$He is bound by $S_n$ = 0.1\,MeV. 
	The contribution of all other sets or combinations of basis states where the two neutrons are in the continuum for a bound $^5$He is negligibly small.
	
	On the other hand, the picture is more intriguing when there is no bound state in the intermediary $^5$He nucleus. The continuum states tend to couple amongst themselves, showing relevant contributions of several configurations despite the presence of a resonance. However, akin to the bound case, all the coupling configurations with {at least} one particle in the resonance basis state {(shown by the red shaded bars in Fig. \ref{fig: basis_contri})} contribute considerably and the total contribution of this third excited state in the continuum basis accounts to about 57\%. {When both particles are in the resonance state, the contribution is about 21\% only, but still the largest.} Going away from the resonance, the contribution starts to decrease owing to the smaller probability of occupation of the given state. Again, the diagonal elements or the basis states with pairing contribute more than others, except for those in which one of the states is a resonance (cf. configuration $|2,3\rangle$).
	
	For completeness, the probability contributions of the basis states were also computed for a lab beam energy of 21\,MeV, {but} with  6 states. The contributions of the dominant states increased to 99.7\% and 61\% {for the bound and unbound case}, respectively. This increase is expected as the lower beam energy does not populate the higher excited states of the system and hence, involves lesser number of couplings between the chosen states.
	The two-neutron cross-sections, at this beam energy of 21\,MeV, comes out to be 6.95 and 0.95\,$\mu$b for the {two cases under consideration}. Comparing these with an unperturbed system ($g=0$), we obtained cross-sections of 6.89 and 0.51\,$\mu$b, respectively. 
	
	Meanwhile, including 8 states for a beam energy of 100\,MeV, which, in principle, means enabling the population of higher lying states, gives us a {much larger} two-neutron transfer cross-section than at 21\,MeV beam energy. This increase can partly be attributed to the higher beam energy providing higher cross-sectional values rather than the inclusion of high lying states, the reason being that the probability coefficients of these higher lying states are too small to result in an increase in the transfer cross-section that is so substantial {for the bound case}. This can be verified from panel a) of Fig. \ref{fig: basis_contri}. {However, the inclusion of higher lying states in the continuum is essential because they contribute significantly to the total transfer process.}
	
	Noteworthy {is} a comparison of cross-sections between the perturbed and unperturbed cases {to check for any enhancement in the cross-sections}. For the bound $S_n$ = 0.1\,MeV arrangement, the cross-sections increase with the introduction of pairing, although this increase is seen to be very small. The increase, however, is the {largest} in the case of the continuum: up to 5 times when the beam energy is 100\,MeV, going from 8.6\,$\mu$b for an unperturbed case to about 44\,$\mu$b for a paired system. Even at the lower beam energy, the increase in the cross-section is almost twice. {This feature, that the increase in cross-section at higher energies, for the continuum, occurs despite the reduced coupling strength of the residual interaction} provides a strong support of pairing enhancement in the region of weakly bound light nuclei. This also sheds light on the role played by the $Q$-value \cite{OV01RPP,KO00EPJ} in the high beam energy regime, and which seems to have little to no effect on two-neutron transfer reactions cross-sections. At the lower beam energy, the $Q$-value role is important as it prevents the population and thus, the contribution of the higher lying energy states of intermediate nucleus, thereby reducing the enhancement.
	
	{We recall that the population probability of the final state in $^6$He is obtained via a coherent sum of all the configurations available in $^5$He. Further, under the case of maximum coherence with equal occupation probabilities for given $N$ states, the enhancement factor (EF), for a heavy nucleus, should be equal to $N$ \cite{Oert91PRC}. Since the higher lying states away from the resonance contribute progressively lesser, then $1\le EF \le N$ in a realistic calculation, should be observed. This is clearly seen in our study for a weakly bound light system near the drip line. In fact,} the very characteristic that even higher lying states in the continuum have significant probability coefficients in the $|i,i\rangle$ configurations coupled with the significant enhancement of two neutron transfer cross-sections due to pairing in a way indicates the ability of a weakly bound neutron rich Borromean ($A$+2) system to exist despite its preceding ($A$+1) nucleus being unbound.

	\section{Conclusions}
	
	We have performed two-neutron transfer reaction calculations in an even ($A$+2) light nuclear system near the dripline, built upon two different hypotheses on the intermediate odd ($A$+1) system. We have compared the more realistic case, corresponding to the Borromean nucleus $^6$He built upon continuum single-particle configurations in $^5$He with a resonance at about 0.7\,MeV, with an artificial case, in which the intermediate system is weakly bound. We observe that for transfer to the continuum, our calculations required a smaller number of states to arrive at results that are consistent for all practical purposes, which otherwise can demand larger model spaces \cite{MN06NPA}.
	
	We show that approaching the driplines, for a weakly bound ($A$+2) system, the use of continuum states of the intermediate odd nucleus is crucial in a two particle transfer process. This inclusion of the continuum helps to understand not only the structure of the final body but also sheds light on the dynamics of the transfer process. The highly correlated (unbound $^5$He) case in which several continuum states are necessary to reproduce the $^6$He ground state, offers many paths through which the sequential two-neutron transfer process can occur with constructive interference. We find that such pairing correlations, which are much more prevalent in the unbound $^5$He case than in the bound case, enhance the transfer cross-sections considerably, depending on the bombarding energy as summarized in Table \ref{T1}. Pairing enhancement originates from the interactive correlations among the particles or excitations causing specific phase relations in a two step process \cite{Oert91PRC}. Enhancement has been observed in pair transfers between heavy semi-magic nuclei \cite{SOS91PLB,DV86PLB}, and could play an important role in the formation and existence of light nuclei near the drip line, especially for Borromean systems. Our results are of direct relevance to the previous and proposed studies of structure and halo properties of $^6$He \cite{KO00EPJ,TT00PRC,Sferrazza}. It will certainly be interesting to examine the scenario with the inclusion of excited states of $^6$He, where one could probably see the effects of BEC- and BCS-like behavior \cite{HSC07PRL} due to changing Cooper pair strength in the final state, but that is left for the future.

	\section*{Acknowedgments}
	This text presents results from research supported by SID funds 2019 (Università degli Studi di Padova, Italy) under project No.~CASA\_SID19\_01. [GS] expresses his gratitude to Antonio M. Moro for discussing the numerical aspects of the work and thanks C. Mondal for feedback with the manuscript.
	
	
	
	\bibliography{gaganbiblio}

\end{document}